\documentclass[%
 preprint, 
 amsmath,amssymb,
 aps, physrev,
]{revtex4-2}

\usepackage{graphicx}
\usepackage{dcolumn}
\usepackage{bm}

\newcommand{\aver}[1]{\left< #1 \right>}

\begin{document}


\title{Phase alignment in a lattice of exciton-polaritonic Bose-Einstein condensates} 

\author{Natalia Kuznetsova, Denis Makarov}
 \email{Contact author: makarov@poi.dvo.ru}
 \affiliation{V.I.Il'ichev Pacific Oceanological Institute
of the Far-Eastern Branch of the Russian
Academy of Sciences, \\ 43 Baltiyskaya St.,
690041, Vladivostok, Russia}
\author{Norayr Asriyan, Andrey Elistratov}%
\affiliation{%
 Center for Fundamental and Applied Research, N. L. Dukhov All-Russia Research Institute of Automatics,  127055, Moscow, Russia
}%


\begin{abstract}
Dynamics of exciton-polariton Bose-Einstein condensate is examined by means of the stochastic Gross-Pitaevskii equation including non-Markovian coupling
to the excitonic reservoir. Attention is concentrated on properties of the condensate lattice created by laser beams providing incoherent pumping
of the reservoir.
It is shown that phase ordering of the lattice depends on temperature. The crossover between the in-phase (``ferromagnetic'') and the checkboard (``antiferromagnetic'')
orders is accompanied by variation of the steady-state condensate density.
Also it is shown that the condensate lattices can retain ordered pattern for temperatures which are much higher
than the critical temperature of a single spot, probably due to suppression of the modulational instability.
\end{abstract}

\maketitle

\tableofcontents

\section{Introduction}
\label{intro}

Interest in exciton-polariton condensates is stimulated by their non-trivial properties.
The first of them is the unusually low effective mass of an exciton-polariton, $10^{-5}$-$10^{-4}$m${_{\mathrm{e}}}$.
It anticipates the possibility of Bose condensation for room temperatures \cite{RuiSu}.
In addition, the exciton-polaritonic condensate is an intrinsically open quantum system,
mainly due to the strong absorption of cavity photons \cite{Gavrilov}.
Compensation of photonic losses requires external pumping.
It can be provided by direct injection of photons into the microresonator and/or incoherent laser pumping
of the exciton reservoir, that ultimately leads to an increase of the condensate density \cite{Kasprzak,Timofeev}.
Thus, the exciton-polariton condensate can be considered as a testing ground for studying non-equilibrium quantum phenomena in open quantum systems.
In particular, the coherent photonic mode can be created through the forced relaxation of condensate polaritons from the reservoir, enabling the development of optical devices such as polaritonic lasers \cite{Zhang-PQE} and optical switches \cite{Chen}.

Another promising field of application of exciton-polaritonic BEC is the simulation of various physical phenomena like the Hawking effect \cite{Jacquet}, Josephson oscillations. Creation of polariton lattices significantly broadens these opportunities. In experiments, they are obtained by superposition of laser beams which produce incoherent pumping of the excitonic reservoir or using periodic external potential. Each condensate spot emits a circularly spreading matter wave which reaches other spots ensuring coupling of the condensates. This allows simulating phase transitions in spin lattices \cite{Zvyag, Berloff-nmat2017, Kalinin_Berloff_NJP2018,  Alyatkin-AFM} and even reproducing phenomena of quasicrystalline physics in polaritonic systems\cite{Alyatkin-AFM}. In addition, lattice-type structures attract much interest in the context of using them for implementations of neural networks\cite{Ballarini2020}, for quantum computations \cite{Ghosh2020}.
Theoretical modeling of such multispot patterns is usually based on the Markovian approximation. However, as it was shown in \cite{PLA20,PLA22,JLTP24},
interaction of the condensate with the excitonic reservoir for temperatures of few tens K is essentially non-Markovian.
Indeed, the spectral width of the excitonic reservoir is commonly much smaller than the splitting caused by exciton-photon coupling.
It means that the decoherence of a condensate polaritonic state interacting with the reservoir can be very slow.
The non-Markovian theory for spatially homogeneous exciton-polariton condensates was developed in \cite{EL18,Quantum}.
Semi-empirical approaches for its generalization onto spatially
inhomogeneous condensates were presented in \cite{PLA22,JLTP24,Bulletin}.
Non-Markovianity implies time-nonlocality caused by memory effects in the dynamics of condensed polaritons.
Memory duration depends on the temperature of the reservoir.
Numerical modeling \cite{PLA22,JLTP24} shows that below 10 K dynamical memory facilitates the onset of a fully coherent condensate with an almost uniform phase of the macroscopic wave function.
As temperature increases, there occurs the crossover from the homogeneous to a fragmented state, accompanied by extensive vortex formation. Above the crossover, decorrelation of vortex/anti-vortex pairs according to the Berezinsky-Kostelitz-Thouless scenario results in emergence of the turbulent regime and destruction of the condensate \cite{BKT_Dagvadorj, BKT_Comaron}.

In the present Letter, we intend to study how this transition is reflected in properties of condensate lattices. Indeed, Bose-Einstein condensates of exciton-polaritons have some specific properties which are not typical for spin lattices. One of the most important is the interaction of exciton-polaritons that can be described by the nonlinear term in the corresponding Gross-Pitaevskii equation.
Due to the impact of nonlinearity, the study of ordered lattices of exciton-polaritonic condensates has an independent value. 


\section{Theory}
\label{sec:Theory}

The exciton-polariton energy spectrum is splitted into two branches. In experiments,
the majority of polaritons belongs to the lower branch,
while the upper one is almost empty. The macroscopic condensate wavefunction of  polaritons $\psi(\mathbf{r},t)$
belonging to the lower branch can be described by the following equation:
\begin{equation}
    i\hbar\frac{\partial\psi}{\partial t} =
    \hat H_0\psi(\mathbf{r},t) +
\hat D\psi(\mathbf{r},t),
\label{sys0}
\end{equation}
where
\begin{equation}
 \hat H_0 = -\frac{\hbar^2}{m_{\text{eff}}}\Delta\psi(\mathbf{r},t)
    + \alpha_{\text{c}}|\psi(\mathbf{r},t)|^2 + \alpha_{\text{r}}\rho_{\text{r}}(\mathbf{r},t),
    \label{Hamilt}
\end{equation}
%
$\alpha_{\text{c}}$ is the coupling strength of condensed excitons,
the constant $\alpha_{\text{r}}$ quantifies energy correction due to  coupling of condensate excitons to the excitonic reservoir (the so-called ``blueshift''),
$\rho_{\text{r}}$ is the excitonic reservoir density, and $\hat D$ is the operator describing non-Hermitian interaction of condensate with the environment.


The non-Hermitian component of interaction with the environment can be represented as a sum of the photonic and excitonic contributions,
\begin{equation}
 \hat D\psi(\mathbf{r},t) = \hat D_{\text{cav}}\psi(\mathbf{r},t)
 + \hat D_{\text{ex}}\psi(\mathbf{r},t).
\end{equation}
The photonic contribution can be fairly described in the Markov approximation that yields
\begin{equation}
\hat D_{\text{cav}}\psi(\mathbf{r},t) = -i\gamma_{\text{cav}}\psi + \eta_{\text{cav}}(\mathbf{r},t),
\end{equation}
where $\eta_{\text{cav}}(\mathbf{r},t)$ is spatially and temporally uncorrelated white noise.
In the truncated Wigner approximation, its autocorrellation function reads \cite{Carusotto}
\begin{equation}
 \aver{\eta_{\text{cav}}^*(\mathbf{r},t')
 \eta_{\text{cav}}(\mathbf{r},t')}
  = \frac{\gamma_{\text{cav}}}{\Delta x\Delta y}\delta(\mathbf{r} - \mathbf{r'})\delta(t-t'),
 \label{cav}
\end{equation}
where $\Delta x$ and $\Delta y$ are grid cell sizes.

Due to the nearly flat exitonic dispersion relation, the excitonic reservoir has a narrow energy spectrum and, consequently, relatively long decoherence time,
especially for low temperatures. It means that
interaction with the excitonic reservoir should be essentially non-Markovian.
In Refs. \cite{PLA22,JLTP24}, the following expression for the operator $\hat D_{\text{ex}}$ is proposed:
\begin{equation}
 \hat D_{\text{ex}}\psi(\mathbf{r}) \simeq \hbar\int\limits_{0}^{t}dt' \Sigma^{\text{R}}(t,t')\psi(\mathbf{r},t')
 + \eta_{\text{ex}}(\mathbf{r},t),
\end{equation}
where $\Sigma^{\text{R}}(t,t')$ is the retarded self-energy function given by the formula
\begin{equation}
  \Sigma^{\text{R}}(t,t') = i\frac{\rho_{\text{r}}^2\alpha_{\text{c}}^2 }{\hbar^2}\frac{e^{-\gamma_{\text{ex}} (t-t')}}{1 + \left[\frac{k_{\text{B}}T}{\hbar}(t-t')\right]^2}\theta(t-t').
  \label{retarded}
\end{equation}
Here $\theta(t)$ is the Heaviside function, and $\gamma_{\text{ex}}$ is the condensate exciton decay rate.
The corresponding expression for the Keldysh component of the self-energy term is
\begin{equation}
  \Sigma^{\text{K}}(t,t') =
  -i\frac{\rho_{\text{r}}^2\alpha_{\text{c}}^2 }{\hbar^2}\frac{e^{-\gamma_{\text{ex}} |t-t'|}}{1 + \left[\frac{k_{\text{B}}T}{\hbar}(t-t')\right]^2}.
  \label{keldysh}
\end{equation}
The Keldysh self-energy is linked to the autocorrelation function of excitonic fluctuations via the relation
\begin{equation}
 \aver{\eta_{\text{ex}}^*(\mathbf{r}, t)\eta_{\text{ex}}(\mathbf{r'},t')} = i\hbar^2\frac{\delta(\mathbf{r},\mathbf{r'})}{\Delta x\Delta y}\Sigma^{\text{K}}(t,t').
 \label{autocorr}
\end{equation}
The density of the excitonic reservoir obeys the equation
\begin{align}
\frac{\partial\rho_{\text{r}}(\mathbf{r},t)}{\partial t} &{=} \frac{1}{\hbar}P_{\text{incoh}}(\mathbf{r},t) {-} \gamma_{\text{exR}}\rho_{\text{r}}(\mathbf{r},t) {-}\frac{2}{\hbar}\text{Im}\left[\psi^*(\mathbf{r},t)\eta(\mathbf{r},t)\right] \nonumber\\
&\quad {-} 2\,\text{Im}\left\{ \psi^*(\mathbf{r},t)
\int dt'\,\Sigma^{\text{R}}(t,t')\psi(\mathbf{r},t')
\right\},
\label{dndt}
\end{align}
where the term $P_{\text{incoh}}(\mathbf{r},t)$ describes the incoherent pumping of the reservoir, $\gamma_{\text{exR}}$ is the reservoir excitons decay rate.

For low temperatures functions $\Sigma^{\text{R}}(t,t')$ and $\Sigma^{\text{K}}(t,t')$ can be approximated by the exponentials
\begin{align}
 \Sigma^{\text{R}}(t,t') &\simeq i\frac{\rho_{\text{r}}^2\alpha_{\text{c}}^2 }{\hbar^2}e^{-\gamma_{\text{eff}}(t-t')}\theta(t-t'),\\
 \Sigma^{\text{K}}(t,t') &\simeq -i\frac{\rho_{\text{r}}^2\alpha_{\text{c}}^2 }{\hbar^2}e^{-\gamma_{\text{eff}}|t-t'|},
 \label{SK}
\end{align}
where $\gamma_{\text{eff}}$ depends on temperature linearly \cite{PLA22}.

The values of memory times $\tau_{\mathrm{eff}}=1/\gamma_{\text{eff}}$ corresponding to typical experimental conditions range from 1 to 10 ps.
In this case the autocorrelation function (\ref{autocorr}) corresponds to the complex-valued stochastic Ornstein-Uhlenbeck process,
\begin{equation*}
    \eta_{\text{ex}} = \frac{\rho_r\alpha_{\text{c}}}{\sqrt{\Delta x\Delta y}}
    \tilde\eta,\quad
    \aver{\tilde\eta(\mathbf{r'}, t')\tilde\eta(\mathbf{r}, t)} =
    \delta(\mathbf{r'} - \mathbf{r})e^{-\gamma_{\text{eff}}|t'-t|}
\end{equation*}
when $\tilde\eta(\mathbf{r}, t)$ is the solution of the following Langevin equation:
\begin{equation}
 \frac{d\tilde\eta(\mathbf{r}, t)}{dt} = -\gamma_{\text{eff}}\tilde\eta(\mathbf{r}, t) + \sqrt{2\gamma_{\text{eff}}}\xi(\mathbf{r}, t).
\end{equation}
Here $\xi(\mathbf{r}, t)$ is a complex white noise with unit variance,
\begin{equation}
 \aver{\xi^*(\mathbf{r}, t)\xi(\mathbf{r'}, t')} = \delta(t-t')\delta(\mathbf{r}-\mathbf{r'}).
\end{equation}
Exponentially decaying memory kernel allows one to reduce the integro-differential evolution equation into an equivalent
time-local equation by introducing an auxiliary wave function,
\begin{equation}
 \phi(\mathbf{r},t) = \psi_0(\mathbf{r})e^{-\gamma_{\text{eff}}t}
 +\int\limits_{0}^t\,dt' e^{-\gamma_{\text{eff}}(t-t')}\psi(\mathbf{r},t'),
 \label{fictious}
\end{equation}
where $\psi_0(\mathbf{r}) = \psi(\mathbf{r},t=0)$. This technique is known as the Markov embedding \cite{DeVega,PLA20,Xiantao,JRLR22}.
The replacement
\begin{equation}
 \int\limits_{0}^t\,dt' e^{-\gamma_{\text{eff}}(t-t')}\psi(\mathbf{r},t') =
 \phi(\mathbf{r},t) - \psi_0(\mathbf{r})e^{-\gamma_{\text{eff}}t},
\end{equation}
reduces integro-differential equations (\ref{sys0}) and (\ref{dndt}) to differential ones,
with addition of the supplementary equation for the auxiliary wave function $\phi$:
 \begin{equation}
 \frac{\partial\phi(\mathbf{r},t)}{\partial t} = \gamma_{\text{eff}}\left[\psi(\mathbf{r},t) - \phi(\mathbf{r},t)\right].
 \label{dmdt}
\end{equation}

\begin{align}
i\hbar \frac{\partial \psi(\mathbf{r},t)}{\partial t}
&=
\hat{H}_0 \psi(\mathbf{r},t)
- \frac{i\hbar \gamma_{\text{cav}}}{2} \psi(\mathbf{r},t)
 \nonumber\\
&\quad +
P_{\text{coh}}(\mathbf{r},t)
+ \eta(\mathbf{r},t)
 \nonumber\\
&\quad +
i \frac{\alpha_{\text{c}}^2 \rho_r^2(\mathbf{r},t)}{\hbar \gamma_{\text{eff}}}
\left[\phi(\mathbf{r},t) - \psi_0(\mathbf{r}) e^{-\gamma_{\text{eff}} t} \right]
,
\label{final_psi}
\end{align}

\begin{equation}
\scalebox{0.93}{$\displaystyle
\begin{aligned}
\frac{\partial \rho_{\text{r}}(\mathbf{r},t)}{\partial t}
&= \frac{1}{\hbar} P_{\text{incoh}}(\mathbf{r})
- \gamma_{\text{exR}} \rho_{\text{r}}(\mathbf{r},t) \\
&\quad - \frac{2}{\hbar} \Im \left\{ \psi^*(\mathbf{r},t) \eta(\mathbf{r},t) \right\} \\
&\quad - \frac{2\alpha_{\text{c}}^2 \rho_r^2(\mathbf{r},t)}{\hbar^2}
\Re \left\{ \psi^*(\mathbf{r},t) \left[\phi(\mathbf{r},t) - \psi_0(\mathbf{r}) e^{-\gamma_{\text{eff}} t} \right] \right\}
\end{aligned}
$}
\label{final_rho}
\end{equation}

Here, equations~(\ref{dmdt})--(\ref{final_rho}) form a closed system of time-local differential equations derived by applying the Markovian embedding technique to the original non-Markovian problem with exponential memory kernel. This approach eliminates the need for explicit evaluation of memory integrals while preserving all essential physical effects, making the system amenable to efficient numerical implementation.

\section{Scheme for numerical simulation}

Numerical simulation was conducted using the the step-wise
scheme based on splitting of the
corresponding propagator into deterministic and stochastic
parts, with sequential their application:
\begin{equation}
    \psi(\mathbf{r}, t + \delta t) = \hat U_{\text{d}}\psi(\mathbf{r},t) + \delta \psi_{\text{noise}}(\mathbf{r},t),
\end{equation}
where $\delta t$ is the integration step,
$\hat U_{\text{d}}$ is the propagator of the `deterministic'' part of \begin{equation}
    i\hbar\frac{\partial\psi}{\partial t} =
    \hat H_0\psi(\mathbf{r},t) +
\hat D\psi(\mathbf{r},t),
\label{sys0b}
\end{equation}
and $\psi_{\text{noise}}$ is a stochastic increment.
Assuming abcense of confinement in the horizontal plane, the computational domain
is limited by using absorbing boundary conditions realized using the well-known concept of perfectly-matched layers (PML).

The deterministic part of the propagator can be described as
\begin{equation*}
    \hat{U}_{\text{d}} = \hat T\left[\exp\left(-\frac{i}{\hbar}\int\limits_{t}^{t+\delta t}\hat H_0\,dt\right)\right],
\end{equation*}
that corresponds to the noiseless equation
\begin{equation}
    i\hbar\frac{\partial \psi}{\partial t} = \hat H_0 \psi
\end{equation}
We utilize the split-step Fourier algorithm for simulation of the deterministic part.
The stochastic increment is given by
\begin{equation}
\begin{aligned}
    \delta \psi_{\text{noise}}(\mathbf{r},t) = \int\limits_{t}^{t+\delta t}
    \left[\eta_{\text{cav}}(\mathbf{r},t)
 + \eta_{\text{ex}}(\mathbf{r},t)\right]\,dt' =\\
 \delta \psi_{\text{cav}} + \delta \psi_{\text{ex}},
 \end{aligned}
\label{dA0}
\end{equation}
Taking into account (\ref{cav}),
 we have
\begin{equation}
    \delta \psi_{\text{cav}}=\int\limits_{t}^{t+\delta t}
    \eta_{\text{cav}}(\mathbf{r},t) \,dt' = \sqrt{\frac{\gamma_{\text{cav}}\delta t}{\Delta z\Delta y}}Y_0,
\end{equation}
where $Y_0$ is a complex-valued stochastic Gaussian variable with unit variance.
The excitonic part of the stochastic increment can be expressed as
\begin{equation}
    \delta\psi_{\text{ex}}(\mathbf{r}, t) = \frac{\alpha_{\text{c}} \rho_{r}}{\sqrt{\Delta x\Delta y}} f(\mathbf{r},t),
\end{equation}
where $f$ is an increment of the normalized Ornstein–Uhlenbeck process:
\begin{equation}
    f(\mathbf{r},t) = \int\limits_t^{t + \delta t} \tilde \eta(\mathbf{r},t')\,dt'.
\end{equation}
Realizations of $\tilde \eta(\mathbf{r},t')$ and $f(\mathbf{r},t)$ are computed by means of the following mappings:
\begin{equation}
\begin{aligned}
    \tilde\eta(\mathbf{r}, t + \delta t) &= \tilde\eta(\mathbf{r},t) + \sqrt{2\gamma_{\text{eff}}}Z_1(\mathbf{r}, t)
    - \gamma_{\text{eff}} \tilde\eta(\mathbf{r}, t)\delta t - \\
    &\quad - \sqrt{2\gamma_{\text{eff}}^3}Z_2({\mathbf{r}, t})
    + \frac{\gamma_{\text{eff}}^2}{2} \tilde\eta(\mathbf{r}, t)\delta t^2,
\end{aligned}
\label{etapsi}
\end{equation}
\begin{equation}
    f(\mathbf{r}, t+\delta t) = \tilde\eta(\mathbf{r},t)\delta t + \sqrt{2\gamma_{\text{eff}}}Z_2(\mathbf{r}, t) - \frac{\gamma_{\text{eff}}}{2} \tilde\eta(\mathbf{r},t)\delta t^2.
    \label{fnpsi}
\end{equation}
The quantities $Z_1(\mathbf{r}, t)$ and $Z_2(\mathbf{r}, t)$ are stochastic integrals of white noise:
\begin{equation}
    Z_1(\mathbf{r},t) = \int\limits_t^{t + \delta t} \xi\,dt,\quad
    Z_2(\mathbf{r},t) = \int\limits_t^{t + \delta t} Z_1\,dt.
\end{equation}
They can be represented as
\begin{equation}
    Z_1 = \sqrt{\delta t}Y_1,\quad
    Z_2 = \delta t^{3/2} \left(\frac{Y_2}{2} + \frac{Y_3}{2\sqrt{3}}\right),
\end{equation}
where $Y_1$, $Y_2$, and $Y_3$ are statistically independent complex Gaussian variables with unit variances.
Adding the stochastic increment $\delta\psi_{\text{noise}}$ to the result of the deterministic propagation, we obtain the numerical solution at the next time step.

\section{Phase alignment of lattices}
\label{sec:Order}

The present section is devoted to the numerical simulation of the condensate dynamics.
Our attention is focused on the case of a lattice constituted of several incoherent laser drives,
 \begin{equation}
  P_{\text{incoh}}(\mathbf{r}) = \gamma_{\text{exR}}\rho_0  \sum_{i,j} w(\mathbf{r-r_{i,j}}).
 \end{equation}
The function $w(\mathbf{r})$ determines the intensity distribution for each spot.
In our model it is taken of the super-Gaussian form,
 \begin{equation}
  w(\mathbf{r-r_{i,j}}) = \exp\left[-\left(\frac{\mathbf{r} - \mathbf{r_{i,j}}}{\sigma_{\text{{r}}}}\right)^{20}
  \right].
 \end{equation}
It corresponds to an almost homogeneous intensity distribution within a circle of radius $ \sigma_{\text{{r}}} $.
In the present Letter the spot centers $\mathbf{r_{i,j}}$ are arranged to form a square lattice of 4 or 9 spots.
The radius of each spot is 2.5 $\mu$m,
and the distance between centers of neighbouring spots is 10 $\mu$m.
The grid has spacing of 0.5$\mu$m that is large enough to satisfy conditions of applicability of the truncated Wigner representation of fluctuations. On the other hand it allows for nearly-circular form
of the pumping spots.
The case of a single spot is also considered.

\begin{figure}[h]
\includegraphics[width=0.8\textwidth]{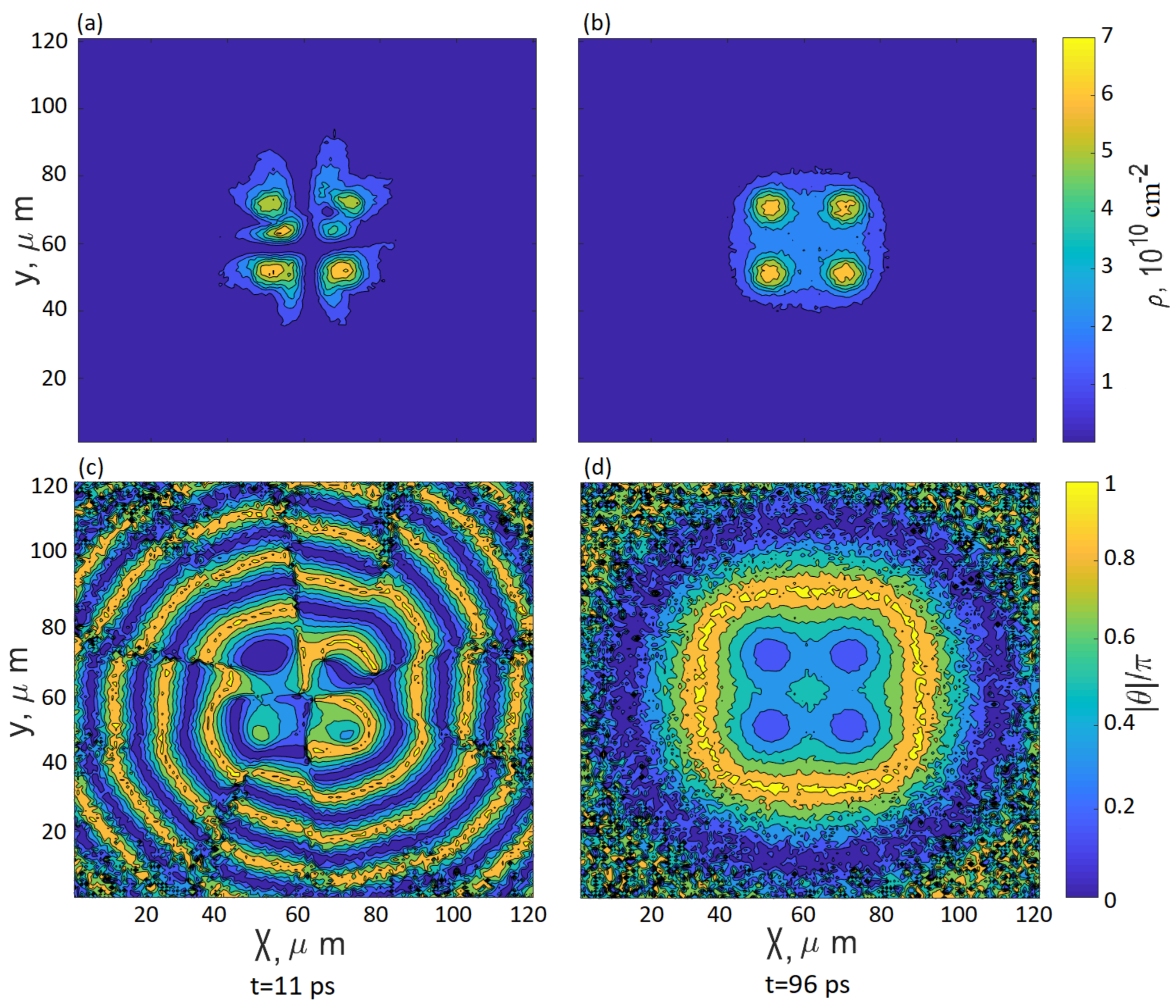}
\caption{Snapshots of spatial density (panels (a) and (b)) and phase (panels (c) and (d)) distributions at different time instants
for temperature 5 K.}
\label{Fig-shot05}
\end{figure}

\begin{figure}[h]
\includegraphics[width=0.8\textwidth]{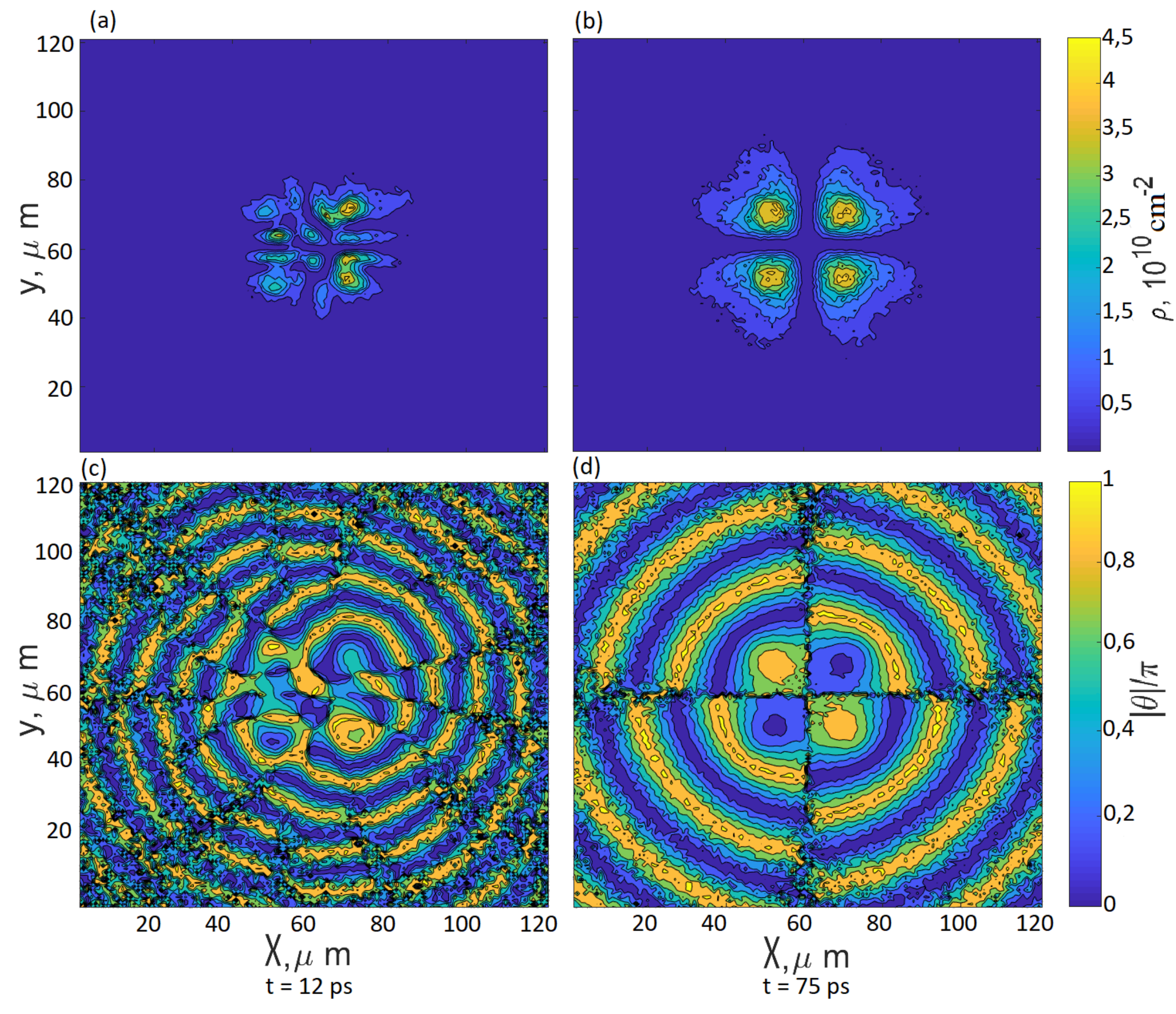}
\caption{Snapshots of spatial density (panels (a) and (b)) and phase (panels (c) and (d)) distributions at different time instants for temperature 20 K.}
\label{Fig-shot20}
\end{figure}

As in our preceding papers, we are interested in the process of spontaneous condensate formation from the very beginning,
therefore we use zero initial conditions:
\begin{equation}
 \psi(t=0) = \phi(t=0) = \rho_{\text{r}}(t=0) = 0.
\end{equation}
In this case the evolution of the system starts from the filling of the excitonic reservoir. This process is accompanied
by generation of fluctuation fields $\eta_{\text{cav}}(\mathbf{r})$ and $\eta_{\text{ex}}(\mathbf{r})$
which serve as seeds for condensate formation.
Spontaneous nucleation of phase coherence domains in vicinities of the pumping spots leads to progressive space-time phase synchronization  of neighbouring areas with these domains.
This process results in flattening spatial phase variations.
It enhances transfer of excitons from the reservoir to the polariton condensate that results in the growth of the average condensate density inside the spots and the formation of a macroscopically coherent states, i.~e. the Bose-Einstein condensate. The rate of coherence emergence strongly depends on the dynamical memory time, which in turn depends on the temperature \cite{PLA22,JLTP24}. Each of the condensate spots emits outflowing waves which can reach other spots providing effective interaction between them. This interaction can eventually result in onset of phase locking and ordered phase configuration of the overall lattice
\cite{Ohadi2016,Berloff-nmat2017}.

The outflowing wave can be approximated as  \cite{Ohadi2016}:
\begin{equation}
 \psi(r) \simeq \sqrt{
 \frac{\rho_0}
 {k_{\text{c}}r }}
 \exp\left(-i k_{\text{w}}r - 2\gamma_{\text{cav}}\tau\right),
\end{equation}
where $\rho_0$ is condensate density inside the pumping spot.
The wavenumber $k_{\text{w}}$ is given by
\begin{equation}
 k_{\text{w}} = \frac{\sqrt{2\mu m_{\text{eff}}}}{\hbar}.
\end{equation}
Here $\mu$ denotes the single-polariton energy,
\begin{equation}
 \mu = \frac{<\psi|\hat H_0|\psi>}{\bar\rho}
\end{equation}
where $\bar\rho$ is the polariton density averaged over area of the pumping spot.
If there are several pumping spots, then the resulting wave function is a superposition of the spots and outflowing waves coming from them.
The phase configuration of the steady-state interference pattern corresponds to the minimum of the total energy \cite{Berloff-nmat2017}.

Figures \ref{Fig-shot05} and \ref{Fig-shot20} show the spatial distributions of condensate density and phase for typical individual realizations of noises $\eta_{\text{cav}}(\mathbf{r},t)$ and $\eta_{\text{ex}}(\mathbf{r},t)$ at $T=5$ K and $T=20$ K respectively.
The results depicted  correspond to a square of 4 pumping spots.
Here it should be mentioned that phase configuration can be identified in an experiment using the homodyne interferometric techique \cite{Alyatkin2020}.

We observe two evolution stages: the ``early'' stage, when distinct condensate spots are just formed,
and the ``late'' stage, when the phase-locked steady state of the lattice is achieved.
The ``early'' stage represents a transient state in course of the crossover from a fully disordered initial state into a symmetric final one.
Formation of order is accompanied by onset of  numerous phase singularities having form of phase knots and corresponding to vortices.
Number of singularities significantly increases as the temperature grows. They destroy the continuous form of the condensate spots, splitting them into several distinct islands.
In course of the subsequent evolution, the singularities are extruded out of the spots, and the corresponding phase singularities transform into specific distortions of the outflowing wavefronts.

After the vortices are removed from the condensate spots, the lattice eventually reaches a steady state with ordered configuration of phases. As it was shown in \cite{Ohadi2016,Berloff-nmat2017}, these configurations can be of either in-phase ``ferromagnetic'' (FM) order, or ''antiferromagnetic'' (AFM) order with alternating phases. The right columns of panels in Figures \ref{Fig-shot05} and \ref{Fig-shot20} depict the ``late'' stages of evolution,
when these stationary states are already achieved. One may see that the final states for $T=5$ K and $T=20$ K correspond to different orders and condensate densities: FM for $T=5$ K with higher density, and AFM for $T=20$ K with lower density. This effect, namely the variation of steady-state density with change of temperature, was earlier demonstraated in numerical simulations \cite{JRLR22,He2024}.

\section{Stability of the condensate lattice}
\label{sec:Stability}

\begin{figure*}
\includegraphics[width=0.9\textwidth]{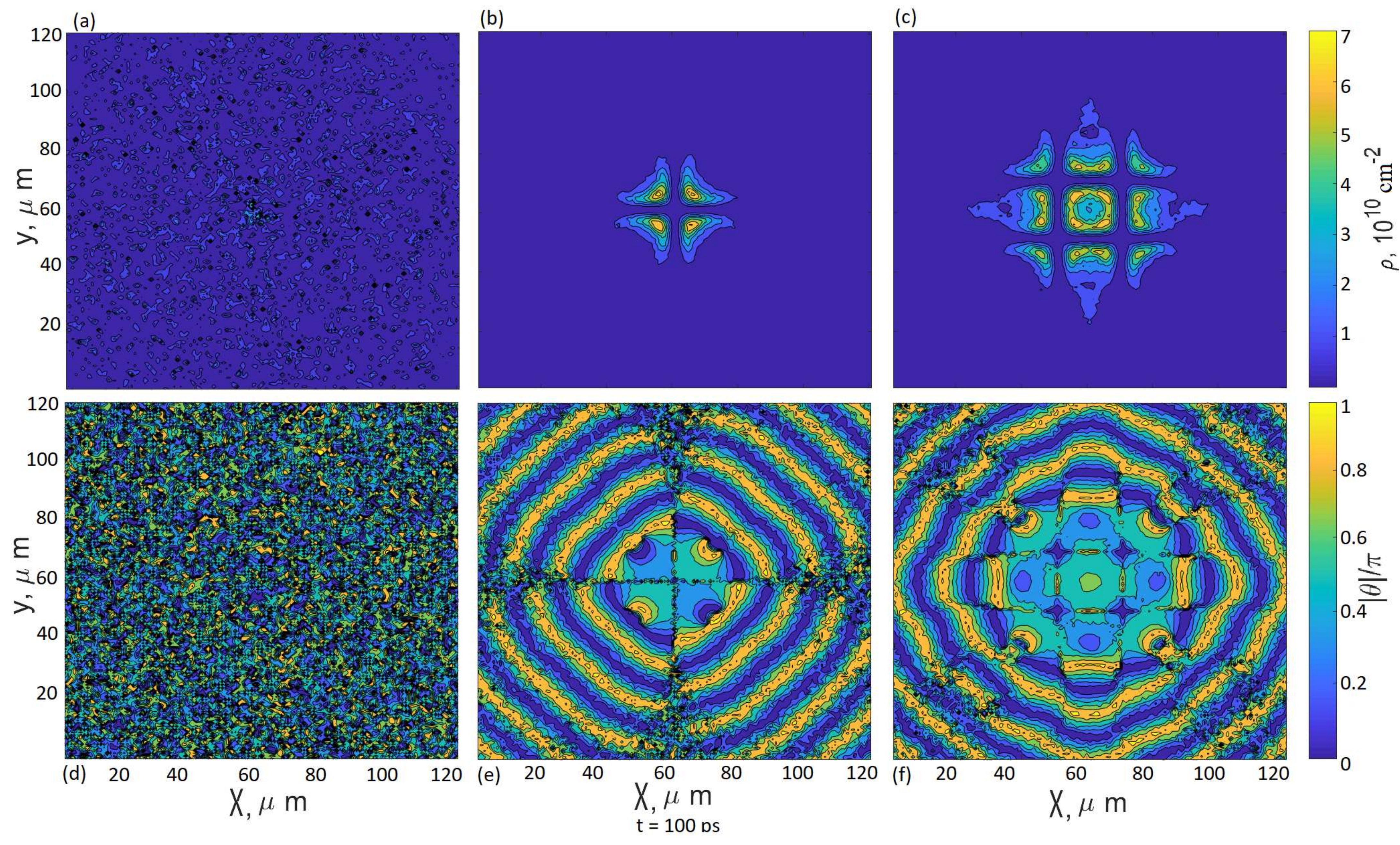}
\caption{Snapshots of spatial density (upper panels) and phase (lower panels) distributions at $t=100$~ps
for temperature 50 K. Panels (a) and (d) correspond to a single condensate spot, panels (b) and (e) correspond to the 2x2 lattice of condensates,
and panels (c) and (f) correspond to the 3x3 lattice of condensates.}
\label{Fig-shot50}
\end{figure*}
The results presented in the previous section show that an increase in temperature leads to a change of the steady-state density and phase ordering
in the lattice, while the phase coherence of the lattice is weakly affected.
On the other hand, phase coherence of a single condensate spot strongly depends on temperature \cite{PLA20,PLA22,JLTP24}.
So, it turns out that coupling of separate condensate spots associated with waves, flowing from different spots, protects the condensate stability against thermal fluctuations.
It is clearly demonstrated in Fig.~\ref{Fig-shot50}, where snapshots of density and phase distributions for $T=50$~K are depicted. There are no signatures
of the condensate onset in the case of the single spot, in marked contrast to the square lattices with $N=4$ and $N=9$ spots.

As long as spatially coherent condensate configurations are characterized by relatively smooth distribution of density, we can measure
coherence by means of the scintillation index defined as
\begin{equation}
 \text{SI} = \frac{\aver{\bar\rho^2}}{\aver{\bar\rho}^2} - 1,
\end{equation}
where the bar denotes averaging over
the area of the spot centered at $x=0$, $y=0$, and the angular brackets denote averaging over ensemble of realizations.

For smooth wave patterns we have $\mathrm{SI}\ll 1$, $\mathrm{SI}\simeq 1$ correspond to statistically saturated intensity fluctuations, $\mathrm{SI}> 1$ indicates the presence of strong stochastic focusing.

\begin{figure}[h]
\resizebox{0.9\textwidth}{!}{%
  \includegraphics{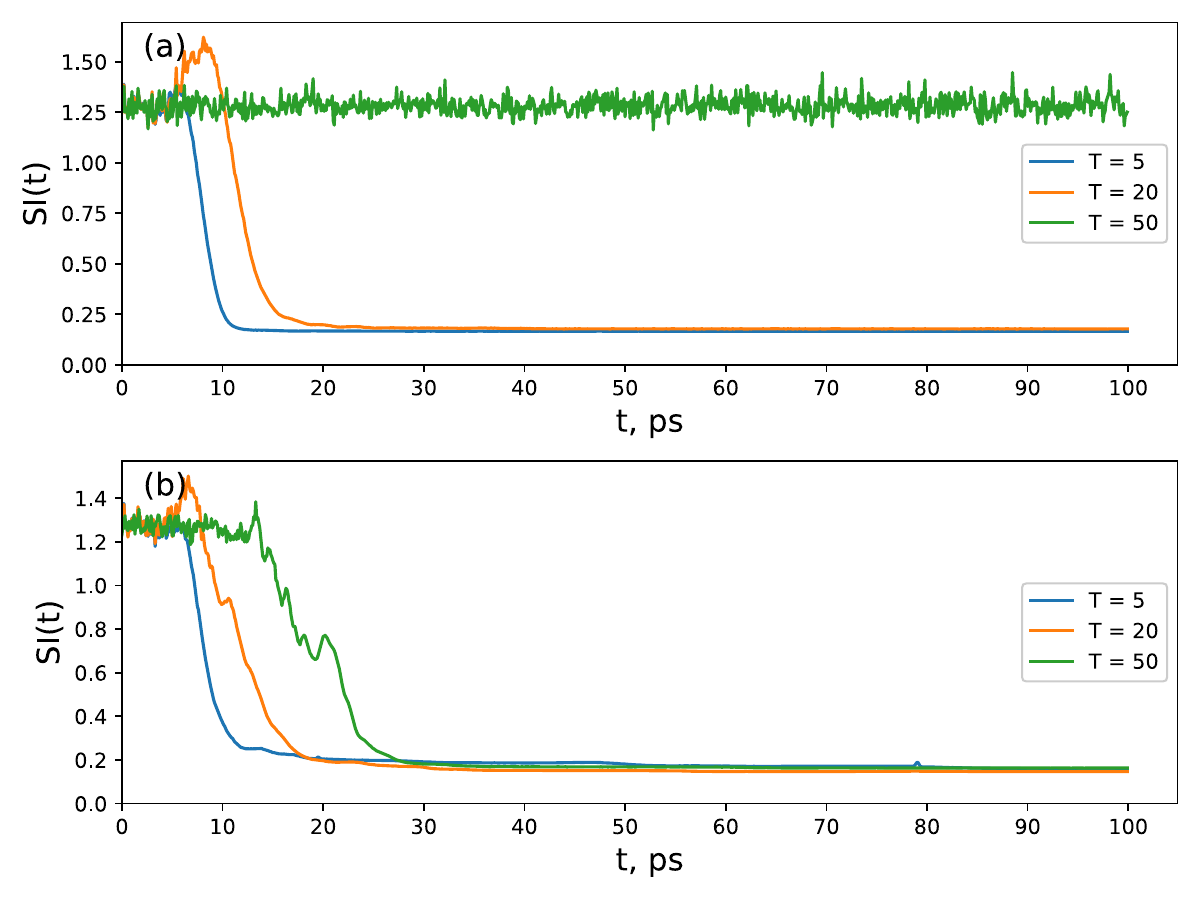}
}
\caption{The scintillation index vs time. Panel (a) corresponds to the single spot, panel (b) corresponds to the 3$\times$3 lattice}
\label{Fig-SI}
\end{figure}

The time dependence of scintillation indices are presented in Fig.~\ref{Fig-SI}.
In the case of the single pumping spot, the scintillation index demonstrates a sharp decrease after the initial transient stage for temperatures of 5 and 20 K,
that indicates the formation of weakly inhomogeneous condensate states.
On the other hand, spatial coherence does not appear at a temperature of 50 K, and SI remains the same as for initial state.
In the case of the 3$\times$3 lattice, the condensate arises for all the temperature values considered, although the initial transient stage duration increases with temperature.


%

In addition, we calculate the first-order coherence function $g^{(1)}(\Delta\mathbf{r},t)$ as a direct indicator of condensate phase coherence. In the present paper, we consider $g^{(1)}(\Delta\mathbf{r},t)$ defined as \cite{RuiSu,JLTP24,Fontaine}
\begin{equation}
    g^{(1)}(\Delta\mathbf{r},t) =
    \frac{|\aver{\psi^*(\mathbf{r_0+\Delta r/2},t)\psi(\mathbf{r_0-\Delta r/2},t)}|}
    {\aver{\sqrt{\rho_{\text{c}}(\mathbf{r_0+\Delta r/2},t)\rho_{\text{c}}(\mathbf{r_0-\Delta r/2},t)}}},
\label{eq:g1}
\end{equation}
with $\rho_{\text{c}}(\mathbf{r},t) = |\psi(\mathbf{r},t)|^2$ being the condensate density.
We use fixed $\mathbf{r_0}=\mathbf{0}$ and average over the two lines: the vertical ($x=0$) and horizontal ($y=0$).

\begin{figure}[h]
\resizebox{0.8\textwidth}{!}{%
  \includegraphics{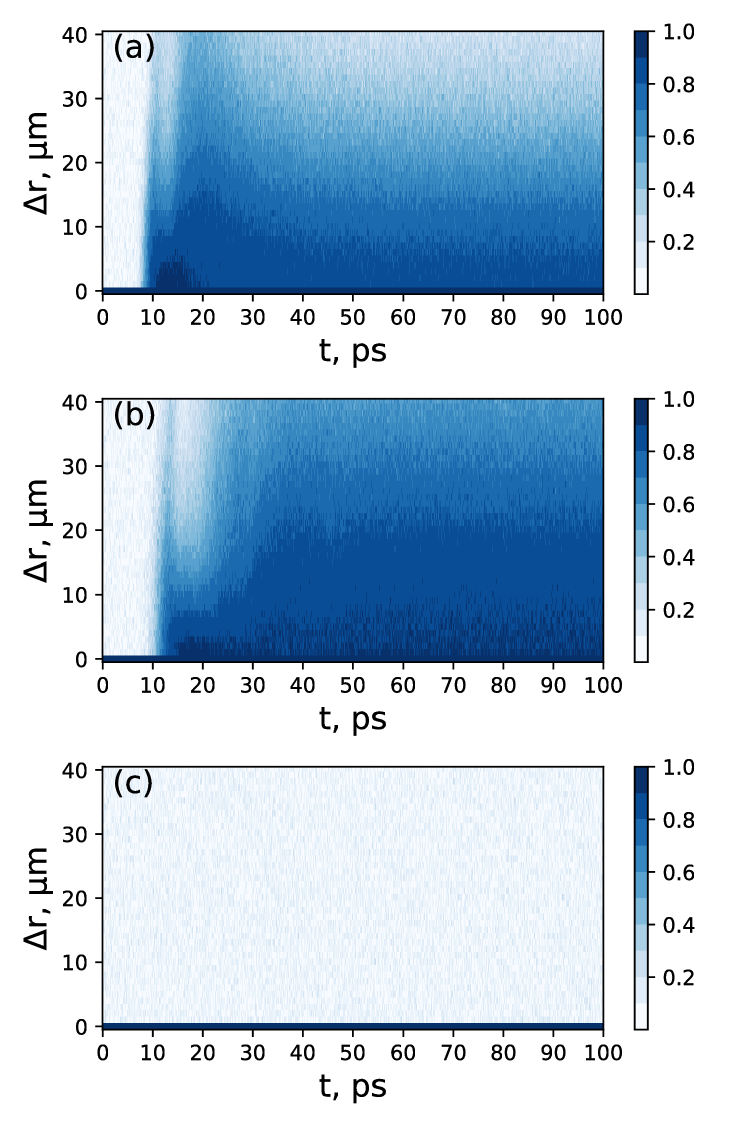}
}
\caption{Function $g^{(1)}$ plotted as function of $\Delta r$ and $t$ for the single pumping spot at {$T=$ (a) 5 K; (b) 20 K; (C) 50 K.}}
\label{Fig-G1-1}
\end{figure}

\begin{figure}[h]
\resizebox{0.8\textwidth}{!}{%
  \includegraphics{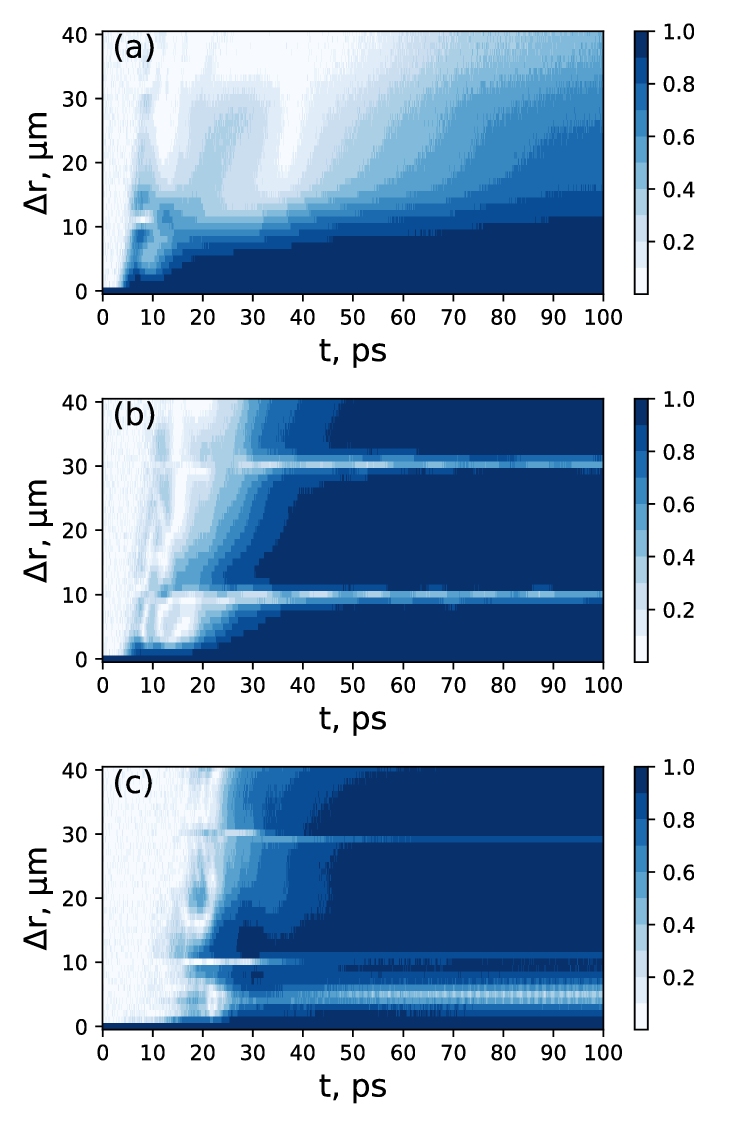}
}
\caption{Function $g^{(1)}$ plotted as function of $\Delta r$ and $t$ for the $3\times 3$ lattice  at $T=$ (a) 5 K; (b) 20 K; (C) 50 K..}
\label{Fig-G1-2}
\end{figure}

Figure \ref{Fig-G1-1} depicts temporal evolution of $g^{(1)}$ for the case of the single pumping spot. For $T=5$ K and $T=20$ K, we observe drastic increase of coherence length occurring near 10 ps and corresponding to
condensate formation. This ``burst'' of coherence lasts a few picoseconds, and then the coherence length strongly decreases. This decrease is associated with expulsion of randomly distributed vortices out of the pumping spot.
After that, a partial restoration of the coherence length occurs. Somewhat surprisingly, the case of $T=20$ K is characterized by slower coherence decay in space than the case of $T=5$ K. Indeed, one should take into account that tails of $g^{(1)}(\mathbf{r})$ are formed by outflowing waves. The case of $T=20$ K corresponds to stronger impact of thermal fluctuations, which facilitates emission of these waves. In the case of $T=50$ K no signature of spatial coherence is observed that indicates the absence of the condensate, as one would expect from the previous results.
For the 3$\times$3 lattice shown in Figure \ref{Fig-G1-2} e,f, long-range coherence is observed for all the temperatures considered (Fig.6). Moreover, the time of coherence onset decreases with increasing temperature, confirming the role of outflowing waves.

There are noticeable light horizontal strips corresponding to coherence drop.
In the case of $T=5$ K and $T=20$ K, they correspond to zones between spots with strongly decreased density. In the case of of $T=50$ K an additional light strip at $\Delta r=5$ $\mu$m appears. It is linked to the distortion of the central spot (see Fig.~\ref{Fig-shot50}(c)).
We suggest that onset of such distortions is related to broadening of condensate spectra with increasing temperature.


To examine the origin of these metamorphoses,
it is reasonable to consider dynamics in the close vicinity of the steady state by fixing condensate and reservoir densities in
(\ref{final_psi}) and (\ref{final_rho}) inside the pumping area:
\begin{equation}
    \rho_{\text{c}}(\mathbf{r},t) \equiv |\psi(\mathbf{r},t)|^2 = \rho_1,\quad
    \rho_{\text{r}}(\mathbf{r},t) = \rho_2
\end{equation}
Also, in the steady state we have
\begin{equation}
    \phi(\mathbf{r},t) = C\psi(\mathbf{r},t),
\end{equation}
where $C$ is some complex-valued constant. Results of numerical simulation show that for the steady states observed
$C$ is purely real and small, $C\ll 1$.
Then one can easily find the relation between $\rho_1$ and $\rho_2$:
\begin{equation}
\rho_2 = \hbar^2\frac{ -\gamma_{\text{exR}} + \sqrt{ \gamma_{\text{exR}}^2 + \dfrac{8C \alpha_{\text{c}}^2}{\hbar^3} \gamma_{\text{exR}}\rho_0 \rho_1 } }{ 4C \alpha_{\text{c}}^2 \rho_1 }.
\end{equation}
Using this relation, we find the effective nonlinearity parameter
\begin{equation}
    \alpha_{\text{eff}} = \frac{d\mu}{d\rho_1} <0,
\end{equation}
i.~e. in the vicinity of the steady state the system is described by
the nonlinear Schr\"odinger equation with attractive nonlinearity.
In this case one can expect onset of the modulational instability \cite{CSF25}. So, we observe suppression of modulational instability
in the lattices.
The phenomenon of condensate stabilization in the lattice was earlier reported in \cite{Baboux} for the condensate lattice formed by etching a planar cavity into a lattice of micropillars. The analysis in \cite{Baboux} associates the stability of the condensate  with the onset of a lattice state with negative effective mass.  In this way symmetric condensate lattices should give rise to steady states with negative effective mass and restored stability. Somewhat surprisingly,
small square lattices also expose such stabilization.
To our knowledge, this effect has never been observed for condensates created by means of incoherent pumping, though a similar effect of vortex ordering in a nonresonantly driven system with optically imprinted polariton lattice has been observed\cite{Alyatkin2024}.

\section{Conclusions}
This Letter discusses spatial patterns arising in a setup with multiple pumping spots in a polariton condensate.
It is shown that the non-Markovian model of condensate evolution predicts the crossover between FM and AFM phase configurations with change of temperature of the reservoir.
Also, we report on surprisingly high stability of the lattice condensate configurations as compared to the case of a single condensate spot.
In particular, we observe well-ordered condensate lattice states for temperature of 50 K, while a single pumping spot does not create condensate under these conditions.

\acknowledgments
Work of N.V. Kuznetsova and D.V. Makarov is supported by
the project No. 124022100072-5 at the Pacific Oceanological Institute of FEB RAS, and by the Foundation of the Advancement of Theoretical Physics and Mathematics ``Basis''.
N.A. Asriyan acknowledges the support by the Russian Science Foundation grant No. 23-42-10010, https://rscf.ru/en/project/23-42-10010/.


\end{document}